\documentclass[aps, pra, a4paper, showpacs, twocolumn, english, 10pt]{revtex4-2}
\usepackage[T1]{fontenc}
\usepackage{babel}
\usepackage{bbm, amsthm, bm, textcomp, nicefrac, geometry, ragged2e}
\usepackage[dvipsnames]{xcolor}
\usepackage{float}
\usepackage[bbgreekl]{mathbbol}
\usepackage{graphicx, epstopdf, color, verbatim, enumitem, ulem}
\usepackage{pbox, array}
\usepackage{mathrsfs}
\usepackage{amssymb}
\usepackage{mathtools}
\usepackage{siunitx}
\usepackage{stackrel}
\usepackage[dvipsnames]{xcolor}
\usepackage[thinlines]{easytable}
\usepackage{amsmath}
\makeatletter

\usepackage[caption=false]{subfig}
\addto\captionsenglish{}

\usepackage{hyperref}
\hypersetup{
    colorlinks = true,
    linkcolor = CadetBlue,
    citecolor =  CadetBlue,
    filecolor = CadetBlue,      
    urlcolor = CadetBlue,
}

\begin{document}
\title{Merging-Based Quantum Repeater}

\author{Maria Flors Mor-Ruiz$^1$}
\thanks{These authors contributed equally to this work.}
\author{Jorge Miguel-Ramiro$^1$}
\thanks{These authors contributed equally to this work.}
\author{Julius Walln\"ofer$^1$}
\author{Tim Coopmans$^{2,3}$}
\author{Wolfgang D\"ur$^1$}

\affiliation{
$^1$Institut f\"ur Theoretische Physik, Universit\"at Innsbruck, Technikerstra{\ss}e 21a, 6020 Innsbruck, Austria\\
$^2$QuTech, Delft University of Technology, Lorentzweg 1, 2628 CJ Delft, The Netherlands\\
$^3$EEMCS, Delft University of Technology, Mekelweg 4, 2628 CD Delft, The Netherlands
}

\begin{abstract}
We introduce an alternative approach for the design of quantum repeaters based on generating entangled states of growing size. The scheme utilizes quantum merging operations, also known as fusion type-I operations, that allow the reintegration and reuse of entanglement. Unlike conventional swapping-based protocols, our method preserves entanglement after failed operations, thereby reducing waiting times, enabling higher rates, and introducing enhanced flexibility in the communication requests. Through proof-of-principle analysis, we demonstrate the advantages of this approach over standard repeater protocols, highlighting its potential for practical quantum communication scenarios.
\end{abstract}

\maketitle

\paragraph*{Introduction.---}
Quantum technologies promise to revolutionize the way we communicate, offering powerful and unique applications within the context of quantum networks \cite{Kimble2008, Wehner2018, Kozlowski2019,Azuma2021}. Some of these applications include quantum teleportation \cite{Bennett_telep}, enhanced sensing \cite{Kessler2014, Eldredge2018,Sekatski2020}, quantum cryptography \cite{Ekert91, Lo2014}, distributed quantum computing \cite{CiracDistributed, Hayashi15,Cacciapuoti2020}, and clock synchronization \cite{ClockSin}. 

A critical component for achieving long-distance quantum communication is the quantum repeater \cite{Briegel_Repeaters, D_r_2016, Van_Meter_2013,Locke2011,Meter_2013,Azuma_2015, Epping_2016, Epping2016_b, Munro_2015,Jones_2016,Li_2019,Wallnofer_2022, Shchukin22, Patil_2022, Kaur2023, MR2023, Romanova24}. This tool enables the distribution of quantum information and entanglement across long distances, overcoming rate scaling bottlenecks, essential for the ultimate goal of developing a viable quantum internet \cite{Kimble2008, Wehner2018}.

Initial proposals of quantum repeaters \cite{Briegel_Repeaters,D_r_2016} rely on the recursive generation of elementary entangled links. These links can be extended to cover larger distances using entanglement swapping techniques \cite{Briegel_Repeaters,D_r_2016},  enabling efficient entanglement distribution rates over such distances. This concept has been extensively analyzed and experimentally validated \cite{Pan98,Liu22,Ning19,Galli2023}. Further extensions and methods have been proposed \cite{Yan_2021}, including first-, second-, and third-generation quantum repeaters. The first \cite{DUR1999} incorporates entanglement purification, while the latter ones \cite{knill1996, Koji2023,Jiang2009, Muralidharan_2016,Borregard2020} rely on quantum error correction and deterministic operations. Limitations of (first-generation) quantum repeaters include the probabilistic nature of entanglement generation and swapping, as well as decoherence effects during qubit storage. Consequently, optimizing strategies to enhance communication rates is essential for enhancing their overall efficiency and practicality.

Here, we propose an alternative approach, denoted as \textit{merging-based} quantum repeater, for designing repeater protocols with improved efficiency and wider flexibility. Our method helps mitigate the impact of the probabilistic nature inherent in some implementations of quantum operations, such as entanglement swapping. In conventional swapping-based repeater setups, the failure of any entanglement swapping operation requires restarting the protocol from scratch, losing all involved entanglement and starting over with the initial elementary entanglement generation. Our method relies on harnessing and recycling entanglement using merging operations rather than entanglement swapping operations, in order to generate multipartite states of growing size that allow for such entanglement recycling. Merging operations are closely related to entanglement swapping in both theoretical principles and practical realization, and have been extensively analyzed theoretically and demonstrated experimentally \cite{PhysRevA.73.022330, Rempe2024Fusion}.   

There are several key features that distinguish our merging-based strategy. First, it has the potential to significantly enhance repeater performance. Since the proposed method preserves most of the established entanglement after a failed merging operation, only the addressing ---or patching--- of small entanglement gaps is required to allow the protocol to continue. These entanglement gaps can be efficiently handled using fixed building blocks, composed of small entanglement structures. This feature leads to improvements in waiting times, particularly for establishing end-to-end connections, in relation to existing strategies. Compared to swapping-based protocols, our method requires quantum memories to store entangled states for extended periods of time, which may reduce the quality of the final states. Nonetheless, this effect is mitigated by the reduced waiting times, resulting in overall enhanced entanglement distribution rates even without including entanglement purification.

Since multipartite entangled structures are built during the process, decisions on the concrete target end-to-end connections can be made at later stages. Hence, our merging-based quantum repeater offers important flexibility advantages. Furthermore, certain parallel or even multipartite entangled connections can be also achieved.

\begin{figure}[h]
    \centering
    \subfloat[]{\label{fig:sb}\includegraphics[width=0.5\columnwidth]{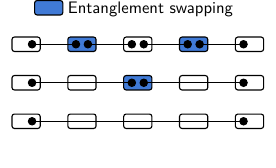}} 
    \hfill
    \subfloat[]{\label{fig:mb}\includegraphics[width=0.5\columnwidth]{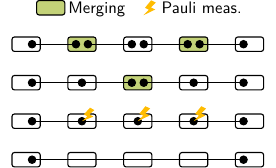}} 
    \caption{Schematic illustration of the steps in a one-dimensional double distance repeater protocol with (a) the existing swapping-based strategy and (b) our merging-based strategy, for a four-segment chain.}
    \label{fig:basic}
\end{figure}

We evaluate the performance of the proposed strategy in the double distance scheme \cite{Briegel_Repeaters, Shchukin2019}, specifically addressing the complex tracking of waiting times and noise. This is achieved by employing a Monte Carlo sampling-based divide-and-conquer algorithm using recursive functions \cite{Shchukin2019, Brand_2020, Julius2024}, and the Noisy Stabilizer Formalism \cite{Mor_NSF}. We demonstrate how improvements in the secret key rate \cite{Abruzzo2013, Brand_2020, Julius2024} can be achieved in relevant regimes. Additionally, we explore and discuss potential enhancements, generalizations, and applications of this approach.

\paragraph*{Setting and general idea.---} 
We address the challenge of establishing quantum connections between two (or more) distant parties, specifically by enabling the distribution of entanglement among them. Losses and noise during the process of entanglement generation and distribution make direct transmission unfeasible. To overcome these limitations, quantum repeaters \cite{Briegel_Repeaters,D_r_2016} serve as fundamental devices that extend the range of quantum communication. Unlike classical communication, where signal amplifiers restore data and compensate for losses, quantum communication requires alternative strategies due to the constraints imposed by the no-cloning theorem \cite{jnielsen2002quantum}, which prohibits the direct amplification of information.

To address this, quantum repeaters segment a communication channel into smaller, manageable distances. Within each segment, entanglement is established locally and then extended across longer segments using entanglement swapping, a process that connects entangled states from adjacent nodes to form bipartite connections over longer distances. These operations, together with entanglement purification to address noise \cite{Briegel_Repeaters,D_r_2016}, enable the reliable distribution of entanglement over large distances, making repeaters indispensable for realizing scalable quantum networks. Existing quantum repeater approaches rely on a series of entanglement swapping operations that generate bipartite entangled states over growing distances. Although effective, these operations are inherently fragile. A failure in any swapping step results in the loss of all previously involved established entanglement, requiring the entire protocol to restart. This introduces significant inefficiencies, particularly as the distance scales.

Our work introduces an alternative framework, the merging-based quantum repeater, which replaces entanglement swapping with merging operations, also known as type-I fusion operations \cite{Browne_Fusion}. Similar to existing first-generation repeater strategies \cite{Briegel_Repeaters, Van_Meter_2013, D_r_2016, Locke2011, Meter_2013, Azuma_2015}, we assume that entanglement generation, entanglement swapping, and merging operations succeed probabilistically. Merging operations iteratively grow larger entangled multipartite states, see Fig.~\ref{fig:mb} for an example in a one-dimensional structure, preserving multipartite entanglement even in the event of a failure. When two multipartite states are successfully merged, a larger multipartite entangled state is obtained. But if a merging operation fails, the two qubits involved are discarded, such that the existing entangled state is not entirely lost. Instead, a localized entanglement gap emerges, requiring only small elementary structures to be regenerated and integrated into the existing multipartite state.

The protocol recursively addresses merging failures by efficiently patching entanglement gaps that arise due to such failures. This approach ensures that most of the entanglement created can be reused, minimizing the need to restart the protocol after each failure event. By preserving the majority of the established entanglement, the merging-based method reduces the effect of a failed operation (recall that the failure of a swapping operation in a existing protocol can destroy all involved entanglement), which in turn lowers the protocol waiting times. When the process is completed, the multipartite state can be locally manipulated to establish connections between any two --or more (multipartite entangled states)-- stations. The choice of the target connection can thus be delayed to the last step of the protocol, which provides the merging-based quantum repeater approach with a flexibility not present in the swapping-based case. 

\paragraph*{Double distance protocol.---}
We illustrate the features of the merging-based quantum repeater by considering the double distance protocol \cite{Briegel_Repeaters, Shchukin2019}, see Fig.~\ref{fig:basic}. We assume a one-dimensional (1D) scheme with $n$ repeater stations arranged in a linear fashion, where each station can establish elementary two-qubit entanglement links with its nearest neighbors. Existing double distance protocols involve iterative entanglement swapping operations performed at every second station, doubling the segment distance of the entangled pair with each step, see Fig.~\ref{fig:sb}. An end-to-end connection can be thus established after $k=\log_2(n-1)$ steps, such that the total distance of the repeater chain is split in $2^k$ segments.

We replace swapping operations with merging ones, sequentially generating growing 1D cluster states \cite{hein_multiparty_2004, hein_entanglement_2006}. We denote the success probabilities of entanglement generation, swapping, and merging operations as $p_{\text{gen}}$, $p_{\text{swap}}$, and $p_{\text{merge}}$, respectively. Given the fundamental features of merging operations \cite{Browne_Fusion}, and their practical similarities with swapping operations \cite{Browne_Fusion}, analogous protocols with similar assumptions can be designed, ensuring a fair performance comparison, therefore, we consider that $p=p_{\text{swap}} = p_{\text{merge}}$. Once the complete entangled structure is formed, in this case a 1D cluster state connecting all parties, see Fig.~\ref{fig:mb}, an additional final step is required, consisting of local Pauli measurements on the intermediate nodes to achieve the desired end-to-end connection.

\begin{figure}[h]
    {\includegraphics[width=\columnwidth]{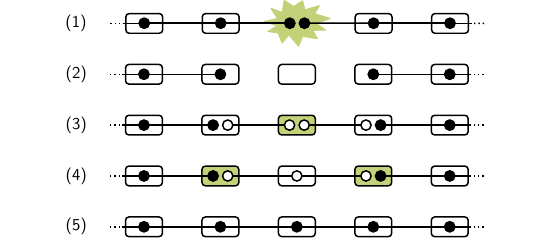}}
    \caption{\label{fig:patch} Patching process. Entanglement gap generated in the event of merging failure and basic building block structure required to patch it. (1) Failed attempt of the merging operation that leaves an entanglement gap (2). (3) Two elementary entanglement structures are generated, denoted with empty circles, and merged into a three-qubit 1D cluster (4), which is then attached via merging operations on the neighbors of the failed event (5).}
\end{figure}

A closer examination of each merging operation reveals a particular behavior. When two $m$-qubit 1D cluster states are successfully merged, a larger 1D cluster of $(2m-1)$ qubits is obtained. If the merging operation fails (step 1 in Fig.~\ref{fig:patch}), the entire entangled state is not lost; instead, an entanglement gap is formed (step 2 in Fig.~\ref{fig:patch}). This gap, equivalent to erasing the unsuccessfully merged qubits, can be efficiently addressed in subsequent protocol steps.

To handle this entanglement gap, only two elementary links need to be regenerated (step 3 in Fig.~\ref{fig:patch}) and merged into a three-qubit cluster state (step 4 in Fig.~\ref{fig:patch}), which is then attached to the original cluster chain (step 5 in Fig.~\ref{fig:patch}).

During this \textit{patching} process, entanglement generation and merging operations used to generate new entanglement and attach it to existing clusters succeed again only probabilistically, consistent with the rest of the protocol. If successful, a larger cluster state is obtained. In case of failure, two relevant situations can arise (step 4 in Fig.~\ref{fig:patch}): (i) only one merging operation fails during the attachment to the existing clusters, such that the entanglement gap is of the same size (two segments in this case) but has been moved to the side where the merging operation has failed; and (ii) both attachment merging operations fail giving rise to a bigger entanglement gap of two more segments (four in this case). We denote these two scenarios as \textit{moving} and \textit{growing} gaps, respectively.

While the patching of moving gaps is attempted recursively, growing gaps require additional considerations. We introduce a \textit{growth limit} $g_l$, which is defined as the maximum number of times we allow the protocol to attempt to grow gaps, before completely restarting it.

\paragraph*{Performance and results.---}
We analyze the performance of our merging-based repeater proposal of the double distance protocol and compare it to the existing swapping-based approach. To do so, we employ the secret key rate as primary figure of merit, widely considered in the literature to analyze repeater performance \cite{Brand_2020, Abruzzo2013, Guha2015, Lucamarini_2018, Rozpdek_2018, Khatri2019, Langenfeld2021, Kamin23, Wallnofer_2022, Julius2024}. We use Monte Carlo simulations \cite{Shchukin2019, Brand_2020,Julius2024} in order to compute the waiting times for both cases, and the Noisy Stabilizer Formalism \cite{Mor_NSF} to evaluate the influence of memory noise during the processes, providing a detailed comparison. 

The secret key rate \cite{Abruzzo2013, Brand_2020, Julius2024}, $S=R r$, which evaluates the overall repeater performance, is given by the product of the raw rate $R$, which describes the speed at which the end-to-end entanglement is distributed, and the asymptotic secret key fraction $r$, which characterizes the quality of the distributed  bipartite entanglement.

The raw rate $R$ \cite{Abruzzo2013, Brand_2020, Julius2024} closely relates to the mean time required for the protocol to complete, taking into account the probabilistic nature of both entanglement generation and swapping and merging operations. In terms of this parameter, our protocol offers a significant advantage, thanks to the ability to patch and reuse entanglement whenever a merging operation fails. 

The secret key fraction $r$ \cite{BENNETT20147}, on the other hand, accounts for noise throughout the process. For both approaches, we consider that the quantum memories are affected by time-dependent single-qubit dephasing noise, whose action on a qubit $v$ is modeled as
\begin{equation}\label{eq:noise}
    \mathcal{E}_v(t)\rho = \left(1-\lambda(t)\right)\rho + \lambda(t)\sigma_z^{(v)}\rho \sigma_z^{(v)},
\end{equation}
where $t$ is the storage duration of a qubit in a quantum memory, and $\lambda(t)=\left(1-e^{t/T}\right)/2$, with $T$ being the dephasing time that characterizes the quality of the quantum memories. 

We make use of Monte Carlo simulations to evaluate the performance of the repeater protocols. For each run we compute: (i) the waiting time needed to establish an end-to-end Bell pair, which allows us to determine the raw rate of each protocol; and (ii) the total storage time for each quantum memory involved in the protocol, for which then we use the Noisy Stabilizer Formalism \cite{Mor_NSF} (see Appendix~\ref{a:noise}) and noise modeled by Eq.~\eqref{eq:noise} to efficiently compute the noisy density matrix of the final end-to-end entangled state, which is needed to evaluate the secret key fraction.

The inherent hierarchical feature of the double distance protocol allows us to split Monte Carlo simulation events and treat them separately, following a recursive structure. We average over multiple rounds and use the sample mean of the quantum bit error rates to estimate the asymptotic secret key rate. We refer to Appendix~\ref{a:montecarlo} for more details.

We verified our numerical simulation by observing good agreement with the analytically computed waiting time on four segments, the smallest case where the merging-based protocol has a different waiting time than the swapping-based protocol. We found this analytical waiting times by writing out the Markov chain of all possible entanglement configurations that can occur. We made use of the fact that the waiting time of simple cases is known, e.g., the waiting time of producing two Bell states in parallel \cite{Shchukin2019}.

\begin{figure}
    \centering
    \includegraphics[width=\columnwidth]{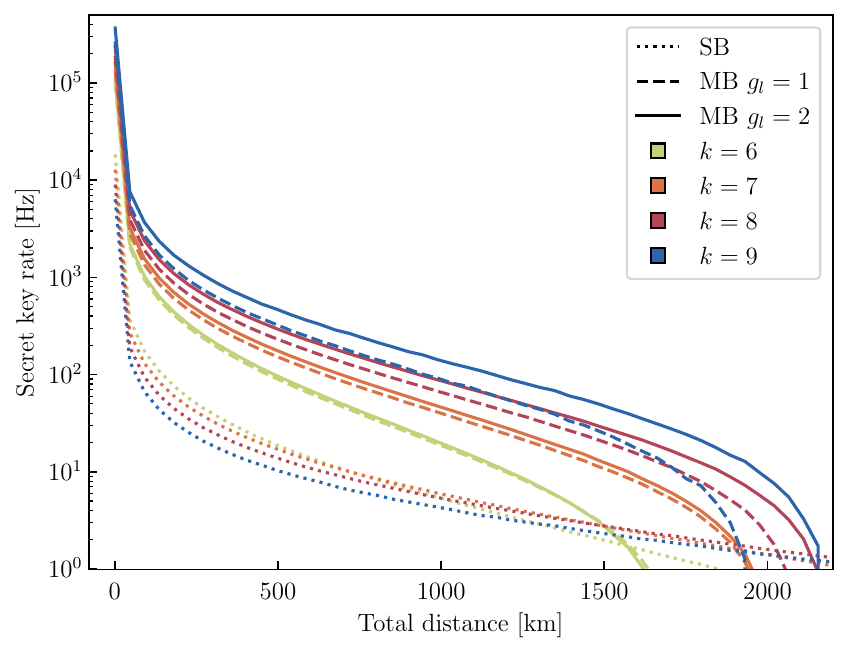}
    \caption{Secret key rate in terms of the total distance, where the number of segments is $2^k$ and the length of each segment is $L_0=L_T/2^k$, where $L_T$ is the total distance, for both the swapping-based (SB) and the merging-based (MB) protocols. The considered MB contains the described limitations to the patch and growth. Both cases are analyzed with quantum memories with \SI[inter-unit-product =$\cdot$]{10}{\second} of dephasing time, a success probability of $p=0.5$ for both the swapping and merging operations, and a generation probability of $p_{\text{gen}}=e^{-L_0/L_{\text{att}}}$, where $L_{\text{att}}=22$ km is the attenuation distance due to transmission losses.}
    \label{fig:SKR}
\end{figure}

In Fig.~\ref{fig:SKR}, we evaluate the secret key rate of the swapping and merging-based repeater approaches as a function of the end-to-end total distance of a linear array of equidistantly arranged repeater stations for different configurations. The results show that the merging-based repeater strategy consistently outperforms the existing swapping-based one, with the improvement becoming more pronounced as more repeater stations are placed along the array, i.e., larger $k$ in Fig.~\ref{fig:SKR}. Furthermore, allowing the patching of larger entanglement gaps (i.e., larger growth limits, $g_l$) further enhances the performance, with the effect being particularly noticeable for chains with more repeater stations. We have also observed that attempts of patching entanglement gaps during the initial steps of the protocol, see Fig.~\ref{fig:mb}, is not beneficial for the overall performance (see also Appendix \ref{a:results}). While patching in the first protocol step is never attempted (since in case of failure at this stage there is no entanglement left), the benefits of patching at following layers in general depend on the protocol parameters.

Despite the overall consistent out-performance of the merging-based approach, the longer coherence times required by our protocol generally result in worse final fidelities compared to swapping-based repeaters (see also Appendix \ref{a:results} for an independent analysis of raw rate and secret key fraction). Therefore, the secret key rate of the merging-based approach drops at shorter distances than the swapping-based one. Nevertheless, the use of entanglement patches to fill gaps partially mitigates this impact of memory decay, since ``fresh'' entanglement, i.e., more recently generated and exposed to less memory decay noise, is introduced at later stages.

In Fig.~\ref{fig:SKR}, for certain distances, the secret key rate is higher for systems with fewer stations. This effect is mainly due to the behavior of the secret key fraction (see Appendix~\ref{a:results}), which is lower for distances on chains with more stations. For a given end-to-end distance and parameters, one can optimize both protocols to determine the precise number of repeater stations that maximize the protocol performance since, on the one hand, more stations are needed to reduce overall waiting times, i.e., to improve the raw rate, but the more memories are involved the more noise affects the entangled state, decreasing the secret key fraction. Nevertheless, note that the dominant parameter is the raw rate, as in general we observe an enhancement of the secret key rate as the number of stations increases in Fig.~\ref{fig:SKR}. For further results and analysis, we refer to Appendix~\ref{a:results}.

These results demonstrate that merging-based quantum repeaters, as presented here, represent a promising alternative in the development of quantum repeaters, offering different possibilities for optimizing their design, performance, and flexibility.

\paragraph*{Further discussion.---} 
While the results presented above highlight the potential benefits of our proposed merging-based repeater strategy, they should be regarded as proof-of-principle findings. In the following, we explore various considerations and generalizations that could further enhance the protocol performance, as well as potential applications and research directions that this alternative approach motivates.

Importantly, the use of temporary multipartite resource states provides greater flexibility, eliminating the need for prior knowledge of specific connection demands. This flexibility enables a wide range of possible entanglement configurations, such as direct connections between any two stations or the creation of multipartite states involving multiple stations, rather than being restricted solely to end-to-end connections.

Observe that so far we consider that an entire entangled structure is generated before intermediate qubits are measured to establish the end-to-end connection. While this provides the flexibility of enabling other connections (or even multipartite ones) at later stages, a stricter focus on end-to-end communication could improve protocol performance. Measuring intermediate qubits earlier, when it is certain they are no longer critical for patching, could reduce detrimental effects of memory noise. However, careful analysis is needed to ensure these qubits are not required for future entanglement gap management. Balancing these factors can enhance efficiency while maintaining robustness.

The underlying principles of our merging-based repeater, the entanglement recycling and patching, could be extended to other areas in quantum networking. This includes applications in distributing multipartite entangled states across more complex topologies that go beyond the one-dimensional repeater chain \cite{Epping_2016, Epping2016_b, Epping_2017, Miguel_Ramiro_2020, Djordjevic_2020, Avis2023, Mor_Influence, Patil_2022}, in multiplexed quantum networks \cite{Munro_2010, Wengerowsky_2018, Asavanant_2019, Haldar2024}, or in supporting simultaneous or parallel entanglement links \cite{Shi_2020, Pirker_2018, Miguel_Ramiro_2023, Freund2024}, enhancing performance and flexibility. 

Overall, the merging-based quantum repeater could offer a practical alternative for short- to medium-term quantum communication networks. With ongoing improvements in quantum memory coherence times \cite{Zhong_2015, Picken_2018, Young_2020, Wang_2021}, its advantages are expected to increase.

\paragraph*{Conclusions.---}
In this work, we have presented an alternative approach for quantum repeaters based on the recursive reuse and recycling of entanglement. Our proposed merging-based quantum repeater protocol offers significant advantages over existing swapping-based strategies, particularly in terms of reducing waiting times and enhancing flexibility when focusing on realistic cases of probabilistic operations. By introducing entanglement patching techniques, our method allows for the preservation of entanglement even after failures of probabilistic repeater operations, significantly reducing the completion times for establishing end-to-end connections.

Our analysis demonstrates that the merging-based approach consistently outperforms common swapping-based strategies across relevant operational regimes. Specifically, it offers advantages in scenarios where repeater stations are relatively close, or where quantum memory coherence times are longer. These characteristics make the approach promising for practical applications in quantum communication structures, where it can thus offer performance and scalability benefits, potentially addressing some of the limitations inherent in current architectures.

Future enhancements to this strategy could focus on addressing memory decoherence effects and integrating other repeater elements. Alternative strategies, inspired by swap-as-soon-as-possible (swap ASAP) protocols, could reduce idle times by initiating merging operations immediately when states become available. Additionally, the implementation of cut-off times could further minimize memory noise and enhance efficiency.

Another promising modification for improvement lies in the integration of entanglement purification protocols, which can mitigate the effects of memory decay and boost fidelity. Techniques specifically tailored for GHZ states and advanced purification methods for cluster states have the potential to enhance performance significantly. Moreover, extending this framework to support the distribution of multipartite and more complex entanglement structures could open new possibilities for quantum networks. Overall, our findings provide an alternative for a more versatile and efficient approach to quantum repeater design, opening a new avenue for advancements in scalable quantum communication.

\paragraph*{Data availability.---} The code we have used for all the results in this work is publicly available at \cite{CODE}. 

\paragraph*{Acknowledgments.---}
M.F.M.R, J.M.R., J.W, and W.D. acknowledge that this research was funded in whole or in part by the Austrian Science Fund (FWF) 10.55776/P36009 and 10.55776/P36010. For open access purposes, the authors have applied a CC BY public copyright license to any author-accepted manuscript version arising from this submission. Also, they acknowledge support by the Austrian Research Promotion Agency (FFG) under Contract Number 914030 (Next Generation EU). T.C. acknowledges the support received through the NWO Quantum Technology program (project number NGF.1582.22.035). All authors thank the Benasque Quantum Information workshop and its organizers, where the core idea for this work was conceived.

\normalem 
\bibliographystyle{apsrev4-2}
\bibliography{refs.bib, refs_appendix}

\renewcommand\appendixname{Appendix}
\appendix

\section{Noise analysis} \label{a:noise}
As outlined in the main text, we consider that all quantum memories are subject to single-qubit time-dependent dephasing noise, modeled by Eq.~\eqref{eq:noise}. Here we present a detailed noise analysis of Bell pairs and cluster states manipulated by swapping and merging operations, and local Pauli measurements, under the action of this type of memory noise.

To study the impact of noise, we employ the Noisy Stabilizer Formalism (NSF) \cite{Mor_NSF}, a method that provides an accurate description of noise evolution. Despite the increased complexity for numerical evaluation of the merging-based repeater protocol, the NSF provides an efficient and precise tool that scales only linearly with the size of the initial state for relevant noise models, as local Pauli-diagonal noise. Moreover, this formalism outputs a complete description of the final noisy state, which is in contrast with other approaches \cite{Gidney2021, Bravyi2019, SimonAnders, PhysRevA.70.052328, cirq} that cannot efficiently evaluate noisy processes, but instead address noise processes through a Monte Carlo methodology, wherein additional gates are inserted randomly. The NSF is build on a class of stabilizer states \cite{stabilizer_gottesman} known as graph states \cite{hein_entanglement_2006, hein_multiparty_2004}, which, as the name indicates, are described by graphs $G = (V, E)$, characterized by a set of vertices $V$ and a set of edges $E$ which represent qubits and the entanglement between them, and where transformations on the state correspond to graphical updates on $G$. Graph states are also defined as the unique $+1$-eigenvalue eigenstate of $K_v = \sigma_x^{(v)}\prod _{(v, w)\in E}\sigma_z^{(w)}$, for all $v\in V$. The NSF independently tracks the graph structure and associated noise operators during protocol execution, using the so-called \textit{update rules}, which are derived from the commutation relations between noise and manipulation operators \cite{Mor_NSF}. After applying all the update rules, the noise operators can be reduced to maps acting only on the output qubits, such that for a relatively small final state one can efficiently reconstruct the final density matrix that describes that state. Note that any stabilizer state is equivalent to a graph state under local Clifford operations \cite{PhysRevA.69.022316}.

We make use of the graph-state basis for our analysis, where the initial Bell pairs are given by $|B\rangle_{\text{ab}} = (|0\rangle_{\text{a}} |+\rangle_{\text{b}} + |1\rangle_{\text{a}} |-\rangle_{\text{b}}) / \sqrt{2}$. Within the NSF framework, all operations are constrained to preserve the graph-state structure. Notably, Pauli measurements include correction operations to ensure that the resulting states remain graph states, although these corrections can performed at a later stage of the protocol due to their commutation properties \cite{MorImperfect, RaussendorfMBQC}.

The merging operation acting on a pair of qubits, the source, $s$, and the target, $t$, qubit is described as 
\begin{equation*}
    \text{MERGE}_{s,\,t} = P_{z,\pm}^{(t)}\text{CNOT}_{s\rightarrow t},
\end{equation*}
where first a CNOT gate between the source and the target is implemented, followed by a measurement in the $\sigma_z$ basis of the target qubit, such that the remaining qubit is the original source one. In the final step of the protocol, intermediate qubits in the 1D cluster are sequentially measured in the $\sigma_y$ basis proceeding from left to right, in order to establish the end-to-end bipartite connection. Similarly, the swapping operation is simulated as a merging operation followed by the measurement in the $\sigma_y$ basis of the source qubit \cite{Mor_Influence, Pirker_2018}.

\begin{figure}[h]
    \centering
    \includegraphics[width=\columnwidth]{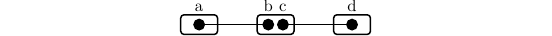}
    \caption{Repeater chain of two segments.}
    \label{fig:example}
\end{figure}

Consider a simple example case involving three stations, four qubits and two segments, where the qubits are labeled as depicted in Fig.~\ref{fig:example}. For simplicity reasons, we assume in the following discussion that operations are deterministic and instantaneous.

At time $t_m$ the merging operation is performed between qubits b and c. Prior to the operation, each qubit undergoes a noise process described by Eq.~\eqref{eq:noise} with $t=t_m$, such that the state is described by $\mathcal{E}_\text{a}(t_m)\mathcal{E}_\text{b}(t_m)\mathcal{E}_\text{c}(t_m)\mathcal{E}_\text{d}(t_m)\rho$, where $\rho=|B\rangle\langle B|_{\text{ab}}\otimes |B\rangle\langle B|_{\text{cd}}$ represents the initial noiseless state. 

When the merging operation is applied, with qubits b and c, designated as the source and target, respectively, the update rule dictates that the noise acting on the target qubit c does not commute with the merging operator. Consequently, this noise requires explicit transformation under the merging operation, while the noise acting on the source qubit b remains unaffected, i.e., $\text{MERGE}_{\text{b},\,\text{c}}\,\sigma_z^{(\text{c})}=\sigma_z^{(\text{b})}\,\text{MERGE}_{\text{b},\,\text{c}}$. Thus, the updated noise maps read
\begin{equation} \label{eq:updated:noise}
    \begin{aligned}
        \mathcal{E}'_\text{a}(t_m)\rho'&= (1-\lambda(t_m))\rho'+\lambda(t_m)\sigma_z^{(\text{a})}\rho' \sigma_z^{(\text{a})}, \\
        \mathcal{E}'_\text{b}(t_m)\rho'&= (1-\lambda(t_m))\rho'+\lambda(t_m)\sigma_z^{(\text{b})}\rho' \sigma_z^{(\text{b})}, \\
        \mathcal{E}'_\text{c}(t_m)\rho'&= (1-\lambda(t_m))\rho'+\lambda(t_m)\sigma_z^{(\text{b})}\rho' \sigma_z^{(\text{b})}, \\
        \mathcal{E}'_\text{d}(t_m)\rho'&= (1-\lambda(t_m))\rho'+\lambda(t_m)\sigma_z^{(\text{d})}\rho' \sigma_z^{(\text{d})},
    \end{aligned}
\end{equation}
where $\rho'=|C_{\text{1D}}\rangle\langle C_{\text{1D}}|_{\text{abd}}$ corresponds to noiseless three-qubit 1D cluster, with $|C_{\text{1D}}\rangle_{\text{abd}}=(|+\rangle_\text{a} |0\rangle_\text{b} |+\rangle_\text{d} + |-\rangle_\text{a} |1\rangle_\text{b} |-\rangle_\text{d}))/\sqrt{2}$. Note that the only updated noise map that differs from the initial configuration is $\mathcal{E}'_\text{c}$,  which adopts the same structure as the noise map for qubit b, such that the combined effect simplifies to $\mathcal{E}'_\text{b}(t)\mathcal{E}'_\text{c}(t)\rho'=\mathcal{E}'_\text{b}(2t)\rho'$. 

From this example there are two main takeaway messages. (i) The noise maps of all the qubits but the target one have the same shape before and after the merging. To account for additional time after the merging, this extra time can simply be incorporated into the pre-existing noise map of each qubit. (ii) For the target qubit memory (i.e., up to the merging operation when its memory ceases due to measurement), the elapsed time can be added to the noise map of the source qubit.

This allows us to compute the total storage times of each of the quantum memories using Monte Carlo simulation (see also Appendix~\ref{a:montecarlo}). Once the final noiseless 1D cluster state is established, dephasing noise maps as described by Eq.~\eqref{eq:updated:noise} are applied, incorporating the computed storage times. Subsequently, measurements in the $\sigma_y$ basis are performed, with noise maps updated using the NSF. Note that, as mentioned before, these measurements are performed left-to-right, which corresponds to the ``side-to-side'' strategy presented for the manipulation of a 1D cluster in \cite{Mor_NSF}.

For the swapping-based case the analysis is the analogous. Storage times for each quantum memory are determined via Monte Carlo simulation, and these times are then integrated into the noise maps of Eq.~\eqref{eq:updated:noise}. Swapped memories are measured in the $\sigma_y$ basis in the order dictated by the swapping operations, with noise maps updated using the NSF. In the double distance protocol, swapping operations follow the ``every-second-qubit''  strategy described in \cite{Mor_NSF}. Note importantly that this property of delaying the application of the noise is not general and depends on the noise model and the operations applied.

We do not explicitly consider noise on the initially generated Bell states. However, the overall impact of such noise on both the merging- and the swapping-based protocols would not significantly alter the comparative results of the secret key rate. This is due to the fact that swapping and merging operations followed by $\sigma_y$ measurements are fundamentally equal.

Note that the choice of the basis and the order of the measurements performed in the intermediate nodes at the latest step of the merging-based protocol could be optimized further in order to minimize the impact of noise in the final Bell pair, as analyzed in \cite{MorImperfect}. While such optimization depends on the storage time of each involved quantum memory, the improvement obtained with an optimal measurement strategy would be noticeable but would not imply a drastic change.

In conclusion, the NSF provides an efficient and reliable framework for numerically evaluating the effects of memory noise in merging-based quantum repeaters, enabling precise comparisons with existing swapping-based protocols.

\section{Monte Carlo simulation} \label{a:montecarlo}
We describe here the details of the code used to numerically evaluate the merging-based repeater performance. The code is written in Python and is publicly available at \cite{CODE}. 

\subsection{Code details}
The Monte Carlo simulation provides samples from the output distribution of waiting times, i.e., how long it takes until the repeater end nodes share an entangled Bell pair. Furthermore, it also tracks how much time each of the qubits in the final cluster state (before the intermediate qubits are measured) has spent in memory to take into account time-dependent memory noise. By considering their statistics from many samples, these two results can be used to calculate the raw rate and the secret key fraction of the protocol, which we use to evaluate the performance of the merging-based repeater.

Our simulation utilizes of the hierarchical structure inherent to the double distance protocol, which allows us to treat the segments at each repeater layer independently because they are not interacting until one performs the merging operation at the next level. This naturally gives rise to a recursive structure, and we can base the code around three core functions that cover all cases.

The first and main function, \texttt{wt}, samples from the waiting time distribution for a specified number of repeater segments. Complementing this, the other two functions, \texttt{wt\_and\_merge\_on\_one\_side} and \texttt{wt\_and\_merge\_on\_two\_sides}, compute the waiting time of producing fresh entanglement and merging the freshly generated entanglement with the existing entanglement. The three functions call each other recursively. The main parameters are the number of segments $s$, the success probability of the generation of an elementary entangled state $p_{\text{gen}}$, the success probability of the merging operation $p$, and the growth limit $g_l$.

Given a repeater chain spanning $s$ segments, the protocol is initiated by calling the function \texttt{wt}. This function splits the total number of segments into two, $s_l$ and $s_r$, and each sub-segment is analyzed independently by calling \texttt{wt} with $s=s_l$ and $s=s_r$. If one sub-segment takes less time than the other, the storage time of the memories holding the earlier segment are adjusted accordingly. For each call, the segments are again split in half and analyzed independently, iterating until $s=1$, which corresponds to an elementary link. At this point, the waiting time for generating the elementary link is sampled from a geometric distribution with bias $p_{\text{gen}}$.  

If large enough $s_l$ and $s_r$ (referring to the fact that due to the patching limitation, the patching of an entanglement gap is only tackled if $s_l, s_r > 4$, as further discussed in Appendix~\ref{a:results}) are successfully created, a merging operation is then attempted with success probability $p$. If it fails (see Fig.~\ref{fig:patch}), an entanglement gap spanning two segments is created. In order to patch it, the function \texttt{wt\_and\_merge\_on\_two\_sides} is called. This function generates a patching block by invoking \texttt{wt}($s=2$), and updates the memory storage times of the pre-existing entanglement ($s_l$ and $s_r$) with the time needed to generate the patch. Two merging operations are then attempted with success probability $p^2$. If one fails, the gap shifts to one side, while keeping the same size, such that \texttt{wt\_and\_merge\_on\_two\_sides} is called again. If both fail, the entanglement gap grows to four segments, and if the tackled segment is large enough and the growth limit, $g_l$ allows it, the patching of the grown entanglement gap is attempted by invoking \texttt{wt\_and\_merge\_on\_two\_sides}. Note that ``large enough'' means that the segment must be larger than $2^{g_{\text{temp}} + 2}$, where $g_{\text{temp}}$ denotes the temporary growth, i.e., how many times the entanglement gap has grown at a certain point in the protocol, due to the patching limitation (further discussed in Appendix~\ref{a:results}). Also, note that for a given $g_l$ the maximum size entanglement gaps to be patched is $2^{g_l+1}$.

The function \texttt{wt\_and\_merge\_on\_one\_side} handles scenarios where entanglement gaps move such that they leave one side without entanglement. It mirrors the structure of \texttt{wt\_and\_merge\_on\_two\_sides} but accounts for single coupling operations. When grown gaps arise, any remaining entanglement in patching blocks is discarded for simplicity, and \texttt{wt\_and\_merge\_on\_two\_sides} is called again. 

This recursive structure ensures scalability, enabling the evaluation of performance for large repeater chains under given characteristics, i.e., noise parameters and success probabilities.

\subsection{Data analysis}
As described in the main text, the secret key rate is calculated as $S=Rr$, where $R$ represents the raw rate and $r$ the secret key fraction. Specifically, the raw rate is the inverse of the average number of rounds needed to successfully distribute a Bell pair between the end nodes, scaled by the appropriate time unit $2L_0/\nu$, where $L_0$ is the elementary distance and $\nu = 2\cdot 10^8$ m/s is the speed of light in optical fiber. The asymptotic secret key fraction of the BB84 quantum key distribution protocol \cite{BENNETT20147} is given by
\begin{equation*}
    r = \max\left(1 - h(\overline{e_x}) - h(\overline{e_z}), 0\right),
\end{equation*}
where $e_x$ and $e_z$ are the Quantum Bit Error Rates (QBERs) \cite{BENNETT20147}
\begin{equation*}
    \begin{aligned}
        e_z &= \langle 01 | \rho |01\rangle + \langle 10 | \rho |10\rangle, \\
        e_x &= \langle +- | \rho |+-\rangle + \langle -+ | \rho |-+\rangle,
    \end{aligned}
\end{equation*}
where $\rho$ corresponds to the final noisy Bell pair and $h(p)$ is the binary entropy function,
\begin{equation*}
    h(p) = -p\log_2(p) - (1-p)\log_2(1-p).
\end{equation*}

Storage times for each quantum memory are used to compute the final noisy density matrix $\rho$. This is achieved using a Python implementation of the NSF for graph states, which is publicly available at \cite{NSF_code}. Since the output noise computation is done in the graph basis, the noisy state is then transformed to the Bell basis by means of a Hadamard gate. Then the QBERs are computed in the standard basis and stored. After enough numerical sampling an averaged waiting time can be estimated, as well as averaged QBERs. With these averaged quantities we can compute the average secret key rate as detailed above.

This approach ensures a comprehensive evaluation of the repeater protocols, accounting for the interplay of waiting times and memory noise in an efficient way.

\section{Further results} \label{a:results}
Here, we provide additional results for a deeper understanding of the influence of memory coherence times and memory decay noise on the performance of the merging-based quantum repeater.

\begin{figure}[b]
    \centering
    \subfloat[]{\label{fig:RR}\includegraphics[width=\columnwidth]{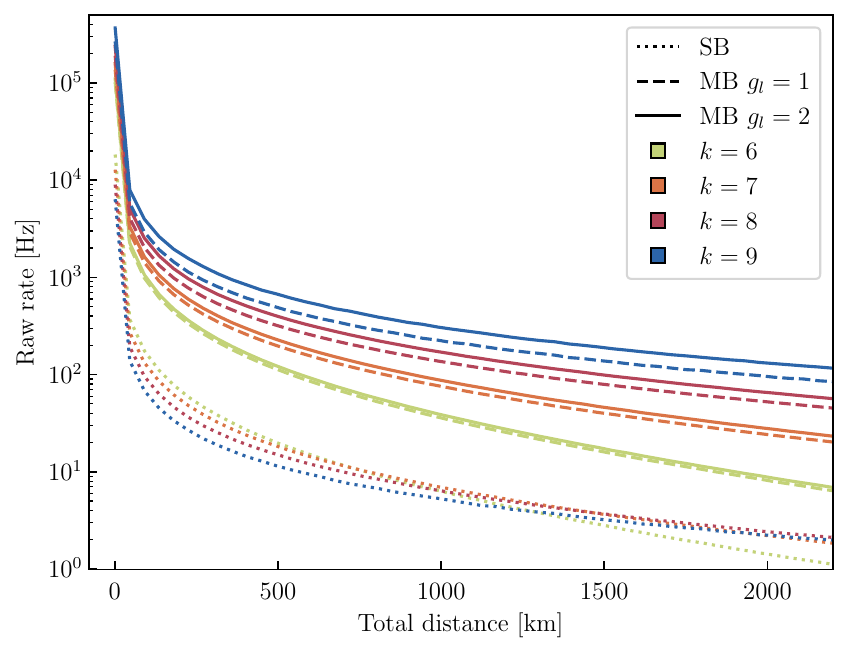}} 
    \hfill
    \subfloat[]{\label{fig:SKF}\includegraphics[width=\columnwidth]{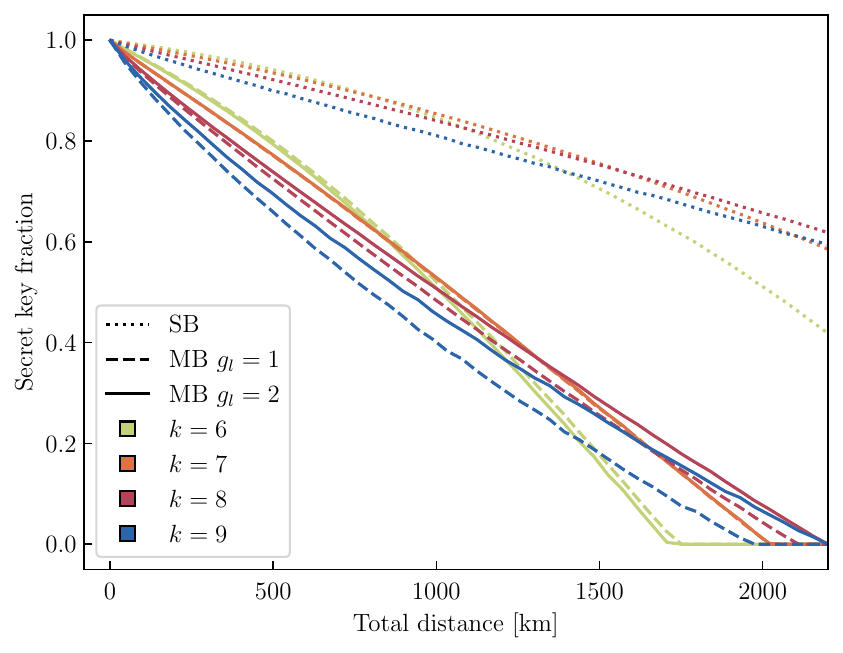}} 
    \caption{Raw rate (a) and secret key fraction (b) in terms of the total distance, where the number of segments is $2^k$ and the length of each segment is $L_0=L_T/2^k$, where $L_T$ is the total distance, for both the swapping-based (SB) and the merging-based (MB) protocols. The considered MB contains the described limitations to the patch and growth. Both cases are analyzed with quantum memories with \SI[inter-unit-product =$\cdot$]{10}{\second} of dephasing time, a success probability of $p=0.5$ for both the swapping and merging operations, and a generation probability of $p_{\text{gen}}=e^{-L_0/L_{\text{att}}}$, where $L_{\text{att}}=22$ km is the attenuation distance due to transmission losses.}
        \label{fig:SKF:RR}
\end{figure}

Fig.~\ref{fig:SKF:RR} illustrates a decomposition of the protocol performance, quantified by the secret key rate $S=R r$, into its two primary components: the raw rate $R$, which captures the average protocol waiting time, and the secret key fraction $r$, which accounts for the effects of noise on the final state.

One can observe that the performance of the merging-based repeater is more sensitive to the additional time-based memory noise that automatically arises when considering longer distances. However, it is important to note that the relationship between the secret key fraction and the fidelity of the final states is not linear, as discussed in \cite{Borregaard2015}. Consequently, the advantage gained in the raw rate has a more substantial influence on the overall performance than the disadvantage associated with the reduced secret key fraction, as demonstrated in the main text.

\begin{figure}[h]
    \centering
    \includegraphics[width=\columnwidth]{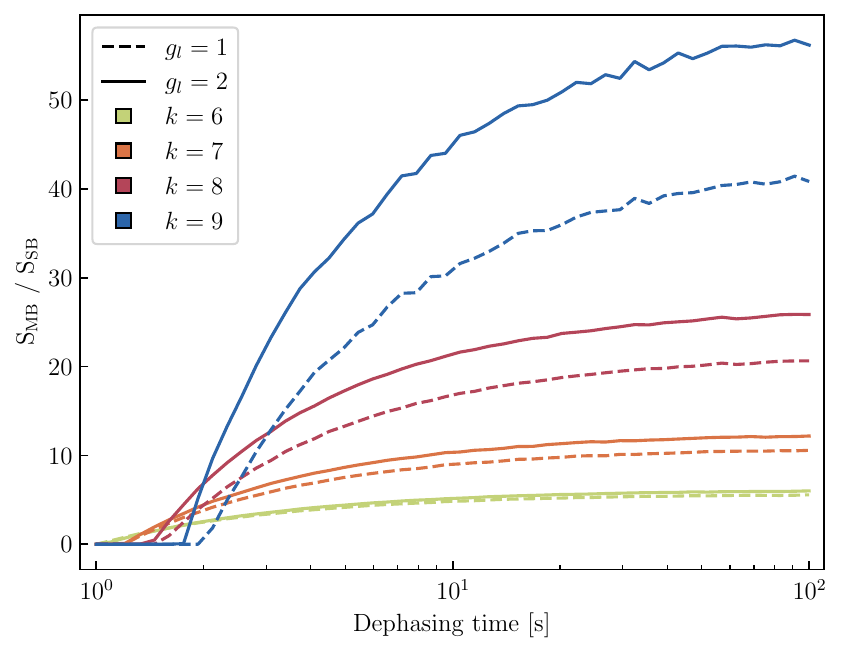}
    \caption{Ratio of improvement on the secret key rate of the merging-based ($S_{\text{MB}}$) over the swapping-based ($S_{\text{SB}}$) protocol in terms of the dephasing time, where the number of segments is $2^k$ and the length of each segment is $L_0=L_T/2^k$, where $L_T=500$ km is the total distance. Two series are plotted for each number of segments corresponding to different growth limits, $g_l$. The considered MB contains the described limitations to the patch and growth. Both cases are analyzed with a success probability of $p=0.5$ for both the swapping and merging operations, and a generation probability of $p_{\text{gen}}=e^{-L_0/L_{\text{att}}}$,, where $L_{\text{att}}=22$ km is the attenuation distance due to transmission losses.}
    \label{fig:ratio}
\end{figure}

The strong influence of the memory quality is further supported by Fig.~\ref{fig:ratio}, in which the improvement ratio of the merging-based repeater over the swapping-based counterpart as a function of the dephasing time of the quantum memories is evaluated. The results demonstrate a consistent advantage across most regimes, with greater benefits observed for longer coherence times. Moreover, in Fig.~\ref{fig:ratio} we observe higher ratios for larger growth limits, $g_l$, i.e., how many times the entangled gap is allowed to grow without restarting the protocol. This highlights the potential of the merging-based approach, particularly in light of the anticipated advancements in quantum memory technologies, which are expected to significantly enhance coherence times in the near future.

\begin{figure}[h]
    \centering
    \includegraphics[width=\columnwidth]{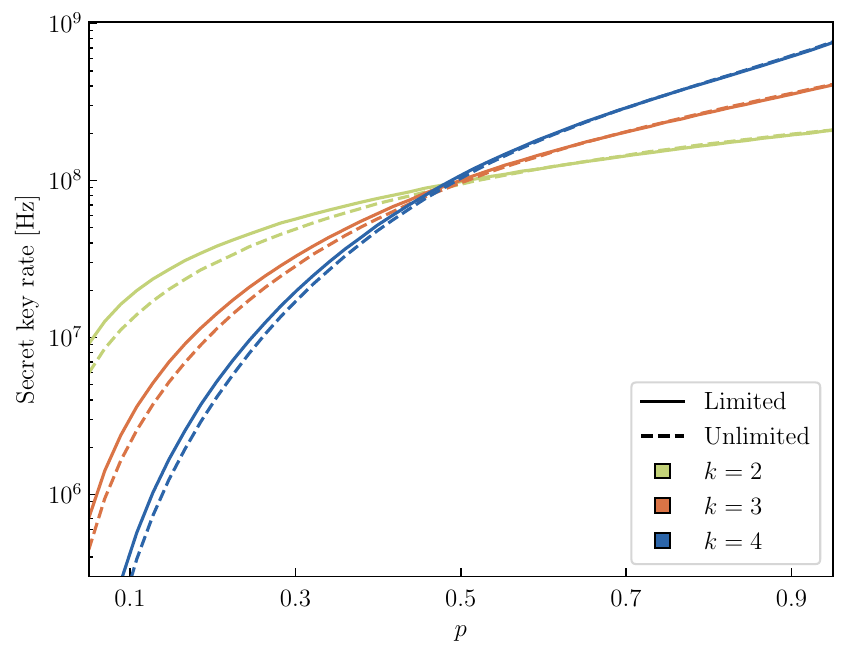}
    \caption{Secret key rate in terms of the success probability of merging operations, where the number of segments is $2^k$ and the length of each segment is $L_0=L_T/2^k$, where $L_T=20$ km is the total distance, for merging-based protocol. Two cases are presented, one with limited patching and one unlimited. Both cases are analyzed with quantum memories with \SI[inter-unit-product =$\cdot$]{10}{\second} of dephasing time and a generation probability of $p_{\text{gen}}=e^{-L_0/L_{\text{att}}}$, where $L_{\text{att}}=22$ km is the attenuation distance due to transmission losses.}
    \label{fig:limit}
\end{figure}

A further detail of the patching strategy is that the remaining states after a merging failure may not be large enough to justify the effort to attempt a patching procedure. Fig.~\ref{fig:limit} illustrates the difference in performance when patching or not patching entanglement gaps during the initial steps of the protocol. We denote these two scenarios as unlimited and limited patching, correspondingly. Note that patching in the first protocol step is never attempted (since in case of failure at this stage there is no entanglement left). The results show that not allowing/limiting the protocol to patch in the first two steps is beneficial for the overall performance. This is because the number of operations required to patch the gap is comparable to or exceeds the effort needed to simply restart the protocol. Consequently, we design our simulations to attempt limited patching when the number of segments under consideration exceeds four. Following the argument for the patching limitation in initial steps, we also consider that the patching of growing gaps will only proceed if the studied number of segments is larger than $2^{g_{\text{temp}}+2}$, where $g_{\text{temp}}$ denotes the temporary growth, i.e., how many times the entanglement gap has grown at a certain point in the protocol.

\end{document}